\documentclass[referee,useAMS,usenatbib]{mn2e}
\def\lsim{\lower.5ex\hbox{$\; \buildrel < \over \sim \;$}}
\def\gsim{\lower.5ex\hbox{$\; \buildrel > \over \sim \;$}}
\def\be{\begin{equation}}
\def\ee{\end{equation}}
\def\eng{{\cal E}}
\def\vel{\vartheta}
\def\ga{\Gamma}
\def\la{\lambda}
\def\md{\dot{\cal M}}

\def\bc{\begin{center}}
\def\ec{\end{center}}
\def\eg{{\it e.g.,}}
\def\etal{{\em et al.}}
\def\ie{{\em i.e.,}}

\usepackage{graphicx}
\title[Effect of flow composition on outflow rates]
{Effect of the flow composition on outflow rates from accretion discs around black holes}
\author[Kumar \etal]
{Rajiv Kumar$^{1}$, Chandra B. Singh$^{2}$, Indranil Chattopadhyay$^{1}$, Sandip K. Chakrabarti$^{3,2}$\\
$^{1}$Aryabhatta Research Institute of Observational Sciences (ARIES), Manora Peak, Nainital-263002, India\\
$^{2}$Indian Centre for Space Physics, Chalantika 43,
Garia Station Rd., Kolkata, 700084, India\\
$^{3}$S.N. Bose National Centre for Basic Sciences, Salt Lake,
Kolkata 700098, India}
\begin{document}
\date{}
\maketitle
\label{firstpage}

\begin{abstract}
We studied the outflow behaviour from accretion discs around black holes taking into account the vertical equilibrium accretion
flow model. The outflow rate is found to depend crucially on flow composition. Our approach is to study the outflow behaviour
as function of inflow around black holes with an equation of state which allows flow to be thermally relativistic close
to black holes and non relativistic far away from black holes. We studied shock ejection model. A pure electron 
positron pair flow never undergoes shock transition while presence of some baryons (common in outflows and jets) 
makes it possible to have standing shock waves in the flow. It can be concluded that the presence of protons is necessary 
for the flow to show the outflow behaviour. The outflow rate is maximum when the flow contains the proton number
density which is 27\% of the electron number density. We conclude that a pure electron-positron jet is unlikely to form.
\end{abstract}

\begin{keywords}
{hydrodynamics, black hole physics, accretion, accretion discs, jets and outflows}
\end{keywords}

\section {Introduction}
One of the spectacular phenomena associated with many classes of astrophysical objects is the formation of outflows and 
jets. Systems which have been associated with winds and jets are active galactic nuclei (AGNs e.g., M87), young stellar 
objects (YSOs e.g., HH 30, HH 34), high mass X-ray binaries (HMXB e.g., SS433, Cyg X-3), black hole X-ray transients (e.g.,
GRS 1915+105, GRO 1655-40) and low mass X-ray binaries (LMXB e.g., Cir X-1). In most of the cases like YSOs, X-ray binaries 
and AGNs, the accretion disc is associated with the production of outflows and jets. In case of YSOs, such as HH 30  
\citep{betal96}, the disc-jet connection has been established. In case of black hole candidates like microquasars and AGNs, 
jets have to originate from the accretion disc, since black holes do not have any atmosphere or hard surface. What is 
even more intriguing, is that the entire accretion disc does not seem to participate in the generation of the outflow 
or jet. VLBI observations of M87 \citep{jbl99} have shown that the jet originates from the immediate vicinity ($\sim 50r_g$) of a
black hole. Moreover, recent detailed observations of various microquasars also show, that the jet states are strongly 
correlated with the spectral states of the associated accretion disc \citep{gfp03,fgr10,rsfp10}.
So the physics of accretion disc is crucial to understand how the jets are launched.

The thin disc, also known as the Shakura-Sunyaev $\alpha$ disc (SS disc) or `standard' disc, was the first successful disc model to explain 
the thermal component of the spectra coming from the black hole candidates \citep{ss73,nt73}. However, the inner boundary 
of the disc was {\it ad hoc}, and the SS disc was unable to explain the hard non-thermal part of 
the spectra. Since the inner boundary condition around a black hole is necessarily transonic, 
models with significant advection gained popularity, and since discs with advection tends to be hotter, 
it became a candidate to explain high energy radiations. Out of all the disc models
in the advective domain, ADAF (advection dominated accretion flow) gained overwhelming popularity \citep{nkh97}, and is a model
characterized by a flow which is mostly subsonic and radiatively inefficient. 
However, it has been shown earlier, that more than one sonic point may exist for inviscid advective flow (Liang \& Thompson, 1980; Chakrabarti, 1989),
and was later confirmed for viscous flow too \citep{c90,c96,lgy99,cd04,bdl08}. It was also shown that,
the ADAF type solutions are a subset of the general advective solutions \citep{c96,lgy99,bdl08,kc13}.  
Interestingly enough, in the multiple critical (or sonic) point \ie MCP domain of the energy-angular 
momentum parameter space, matter passing through the outer sonic point may undergo centrifugal barrier 
induced shock transition \citep{f87,c89,c96,bdl08,kc13}. A shocked accretion disc solution has various advantages.
Numerical simulations established that, the unbalanced thermal gradient term along the vertical 
direction in the post-shock region of the disc, drives bipolar outflows \citep{mlc94,mrc96}. In a model solution starting with
sub-Keplerian (advective) and Keplerian matter, \citet{ct95} showed that the soft, hard and the intermediate spectral
states can be satisfactorily explained. The post-shock disc inverse-Comptonizes the intercepted soft photons to produce the
hard power-law tail and produces the so-called hard spectral state, while absence of or weakened post-shock disc produces essentially
the soft spectral state. These assertions are vindicated by observations \citep{shms01,shs02,sds07}, and also by numerical simulations
\citep{gc13}. Moreover, oscillating and transient shocks explain the low-frequency quasi periodic oscillation (QPO) and its evolution 
during spectral state transitions \citep{ndmc12}.  

Various outflow models exist in the literature \citep{bp82,pc90,c05} which concentrate on collimation and acceleration 
of outflows. However, very few addressed the most important issue of the outflow rate as a function of the inflow parameters. 
\citet{c99} attempted to estimate the outflow rate by considering simple hydrodynamical equations such as those 
applicable to conical inflows and outflows. The work of \citet{c99} was later extended to rotating inviscid 
flows \citep{dcnc01,sc11a,sc11b,sc12}, as well as the viscid flows \citep{cd07,dc08,kc13}. 

Majority of these works mentioned above, however, used a constant polytropic index and did not consider the variation 
of it from a large distance to the black hole and from the disc phase to the jet phase. Realistically, at large distances away from the central object, 
matter should be cold, but as it dives into the central object its thermal and kinetic energy increases. A flow with an equation of 
state described by a fixed $\Gamma$ (adiabatic index) need not describe the behaviour of accreting matter 
properly from infinity to the horizon, because for ultra-relativistic non-degenerate gas $\Gamma=4/3$, 
and that for non-relativistic one, it is $\Gamma=5/3$. $\Gamma$ takes intermediate values
for trans-relativistic thermal energy. The equation of state (hereafter, EoS) of a single species relativistic gas was computed by
\citet{c38,s57}, which gave the correct description of the flow in all temperature range, and we call the EoS as RP
(relativistically perfect). However, RP is a combination of modified Bessel's functions and which makes its usage 
non-trivial and fairly expensive from computational point of view. Hence many workers proposed approximate EoS for single species
flow, which are close to RP and yet easy to use \citep{t48,m71,rcc06}. For the very first time Blumenthal \& Mathews (1976,
hereafter BM76), used the EoS of
\citet{m71} and solved for radial accretion in general relativity. Following the suggestions of \citet{bm76}, 
\citet{c08,cr09} proposed an EoS of multi-species flow composed of electrons, protons and positrons, although
EoS for electron-proton flow was suggested earlier \citep{f87}. \citet{c08,cr09} showed in a general relativistic
investigation that, a purely electron-positron adiabatic flow is the slowest, thermally least relativistic flow, 
compared to flows containing protons and leptons, and consequently, electron-positron flows
do not show accretion shocks \citep{c08,cc11}. It was also shown that electron-proton
(\ie equal number of electrons \& protons) 
flow is not the
most relativistic case. However, when the proton number density is $\sim 20\%$ of the electron number density, 
the flow  becomes most relativistic. Therefore, an estimation of the outflow rate from the inflow parameters using an
equation of state of a fully ionized flow would be an important contribution. In this paper, we precisely do this.
The plan of our paper is the following: in the next Section, we present the assumptions and equations governing inflow and outflow 
from accretion discs around black holes. In Section 3, we present a detailed analysis of critical point behaviour and solution
of accreting flow. We then present our results exhibiting the dependence of outflow rates on 
flow composition, energy and angular momentum of the accretion disc. Finally, in Section 4, 
we carry out discussion and make concluding remarks.
 
\section {Assumptions and Equations}

We consider a fully ionized, adiabatic, rotating and accreting axisymmetric disc around a Schwarzschild black hole. 
For mathematical simplicity, space-time around the black hole is described by the Paczy\'nski-Wiita (hereafter PW)
pseudo-Newtonian potential \citep{pw80}. We have used PW potential in order to simplify our calculations
while retaining all the essential qualitative features of strong gravity. Moreover, it is easier to incorporate
more complicated physics in pseudo-Newtonian scheme. Earlier investigations of disc-jet system had been to employ
general relativity and fixed $\Gamma$ EoS, or general relativity with relativistic EoS, or
PW approach with fixed $\Gamma$ EoS. So far, the approach with PW potential which is coupled with
a variable $\Gamma$ EoS was not attempted and ours is the first to do so.       

\subsection{Equation of state}

Following \citet{c08,cr09}, we consider flows which are composed of electrons ($e^{-}$), positrons ($e^{+}$) and
protons ($p^{+}$) of varying proportions, but always maintaining the overall charge neutrality.
The number density is given by,
\be
n = {\Sigma_{i}}n_{i} = n_{e^{-}} + n_{e^{+}} + n_{p^{+}},    
\label{fnd.eq}
\ee

where, $n_{e^{-}}$, $n_{e^{+}}$ and $n_{p^{+}}$ are the electron, positron and proton number densities, respectively. 
Charge neutrality condition over a reasonable volume element demands that 
\be
 n_{e^{-}} = n_{e^{+}} + n_{p^{+}},
\label{cnu.eq}
\ee
which implies, 
\be
n = 2n_{e^{-}}, ~\mbox{and}~  n_{e^{+}} = n_{e^{-}}(1 - \xi),
\label{icnu.eq}
\ee
where $\xi = n_{p^{+}}/n_{e^{-}}$ is the relative proportion of protons. The mass density is given by
\be
\rho =  {\Sigma_{i}}n_{i}m_{i} =  n_{e^{-}}m_{e^{-}}\left[2 - {\xi(1 - 1/{\eta})}\right],
\label{fde.eq}
\ee
where, $\eta = m_{e^{-}}/ m_{p^{+}}$, and $m_{e^{-}}$ and $m_{p^{+}}$ are the electron and proton masses, respectively. 
For single temperature flows, the isotropic pressure is given by
\be
p = \Sigma_{i}p_{i} = 2n_{e^{-}}kT=2n_{e^-}m_{e^{-}}c^2\Theta.
\label{fp.eq}
\ee

The EoS for multi-species flow is \citep{c08,cr09}
\be
\bar e = \Sigma_{i}e_{i} = \Sigma \left[n_{i}m_{i}c^{2} + p_{i}\left(\frac {9p_{i} + 3n_{i}m_{i}c^2}{3p_{i} + 2n_{i}m_{i}c^2}
\right)
\right].
\label{eed.eq}
\ee
The non-dimensional temperature is defined with respect to the electron rest mass energy, $\Theta = kT/(m_{e^{-}}c^2)$.
Using equations (\ref{fnd.eq})-(\ref{fp.eq}), the expression of the energy density in equation (\ref{eed.eq}) simplifies to, 
\be
\bar e = n_{e^{-}}m_{e^{-}}c^{2}f = \rho_{e^{-}}c^2f = \frac{\rho c^2 f}{[2 - \xi(1 - 1/{\eta})]},
\label{seed.eq}
\ee
where,
$$
f = (2-\xi) \left[1 + \Theta \left(\frac {9\Theta + 3}{3\Theta + 2}\right)\right] + \xi \left[\frac{1}{\eta} + \Theta
\left(\frac {9\Theta + 3/\eta}{3\Theta + 2/\eta}\right)\right].
$$
The enthalpy is given by, 
\be
h = \frac{(\bar e + p)}{\rho}  =  \frac{fc^2}{K} + \frac{2c^2\Theta}{K},
\label{enth.eq}   
\ee
where,
$
K = [2 - {\xi(1 - 1/{\eta})}].
$

The expression of the polytropic index is given by,
\be
N = \frac{1}{2} \frac{df}{d\Theta}.
\label{pol.eq}
\ee
The adiabatic index is
\be
\Gamma = 1 + \frac {1}{N}.
\label{adi.eq}
\ee

\subsection {Equations of motion}
The length, time and velocity scales are measured in units of 
$r_g=2GM_{BH}/c^2$, $r_g/c=2GM_{BH}/c^3$ and $c$, respectively, where, $r_g$ is the Schwarzschild radius, $M_{BH}$ is the mass
of the black hole, $G$ is the gravitational constant, and $c$ is the velocity of light, this amounts to $2G=M_{BH}=c=1$.
Hereafter, all equations are written using the above unit system. It is to be noted though, we have
retained the same symbols for the thermodynamic quantities
like $\rho$, $p$ etc, in the dimensionless form too.
\subsubsection {Inflow}
In our study we neglect the non-conservative processes and magnetic fields. We consider the steady and radial flow assumptions.

In this case, the radial momentum equation is given by,
\be
\vel \frac{d\vel}{dx} + \frac{1}{\rho} \frac{dp}{dx} - {\frac {\lambda^2}{x^3}}  + \frac{1}{2(x-1)^2} = 0.
\label{rme.eq}
\ee
Here, $\lambda$, $\vel$, and $x$ are the specific angular momentum, the infall velocity, and the radial coordinate
in the units described above. The integral form of the eq.(\ref{rme.eq}) gives the Bernoulli parameter
and is written as
\be
\eng=\frac{\vel^2}{2}+ h +\frac{\lambda^2}{2x^2}-\frac{1}{2(x-1)}
\label{Bp.eq}
\ee
and is also called as specific energy of the flow. 

The mass conservation equation is given by,
\be
\dot{M}_{in}=2\pi \Sigma \vel x,
\label{mf.eq}
\ee
where $\Sigma=2\rho H$ is the vertically integrated surface density of the flow. 

The disc matter is in hydrostatic equilibrium in the vertical direction, and the half height is given by \citep{c89}
\be
H=2\sqrt{\frac{\Theta x}{K}}(x-1) .
\label{hh.eq}
\ee
The entropy generation equation or the first law of thermodynamics is given by,
\be
\frac{de}{dx} - \frac{p} {\rho^{2}} \frac{d\rho}{dx} = 0,
\label{cee.eq}
\ee
here, $e={\bar{e}}/\rho=f/K$, where $c=1$.
Integrating eq.(\ref{cee.eq}), and using the definition of $e$, $\rho$ (eq. \ref{fde.eq}),
and $p$ (eq. \ref{fp.eq}), we get 
\be
\rho={\cal{K}} ~\mbox{exp}(k_3)~ \Theta^{3/2}(3\Theta+2)^{k_1}(3\Theta+2/\eta)^{k_2},
\label{aee.eq}
\ee
where, $k_1=3(2-\xi)/4, k_2=3\xi/4$, and $k_3=(f-K)/(2\Theta)$.
This is the adiabatic equation of state for multispecies flows and ${\cal{K}}$ is the constant of entropy.
Using equations (\ref{mf.eq}) and (\ref{aee.eq}), we can define entropy accretion rate ($\md$) as
\be
\md = \frac{\dot{M}_{in}}{4\pi \cal{K}}= \vel H x ~\mbox{exp}(k_3)~ \Theta^{3/2}(3\Theta+2)^{k_1}(3\Theta+2/\eta)^{k_2},
\label{enar.eq}
\ee
here, $\md$ is also constant for inviscid multispecies relativistic flows.

Using equations (\ref{mf.eq}) and (\ref{hh.eq}) in equation (\ref{cee.eq}), we get
\be
\frac{d\Theta}{dx} = -\frac{2\Theta}{2N+1} \left[\frac{1}{\vel} \frac{d\vel}{dx} + 
\frac{5x-3}{2x(x-1)}\right].
\label{dth.eq}
\ee
Using equations (\ref{enth.eq}) and (\ref{dth.eq}) in equation (\ref{rme.eq}), we get  
\be
\frac{d\vel}{dx} = \frac {a^2 [{\frac{2N}{2N+1}} {\frac{5x-3}{2x(x-1)}}] + 
{\frac {\lambda^2}{x^3}} - {\frac {1}{2(x-1)^2}}} {\vel - {\frac{a^2} { \vel}} [{\frac {2N} 
{2N+1}]}},  
\label{dv.eq}
\ee
where, $a$ is the adiabatic sound speed given by, $a^{2} = 2 \Theta \Gamma/K$.

\subsubsection{Critical-point conditions}
Gravity, by the very act of pulling matter towards the gravitating centre, primarily causes the increase
of the infall speed of the matter, but as a secondary effect, also compresses the in flowing matter and thereby increasing
its temperature, and consequently, the sound speed.
This causes the flow velocity to cross the local sound speed at a finite distance, which is termed as the sonic point
or the critical point of the flow. If the matter is rotating, then the centrifugal force opposes the effect of gravity.
Since the centrifugal force is $\propto x^{-3}$ and is axisymmetric, it breaks the spherical symmetry of gravity,
causing the formation of multiple sonic points in a significant range of ${\cal E}-\lambda$ parameter space
\citep{lt80,c89}.
At the sonic point $d\vel/dx \rightarrow 0/0$, which gives us the critical or sonic point conditions,
\be
a_{c}^2 = (\Gamma_c+1) \left( \frac{5x_c-3}{x_c(x_c-1)}\right) \left( \frac{1}{2(x_c-1)^2}-\frac{\lambda_c^2}{x_c^3} \right),
\label{nc.eq}
\ee
and
\be
\vel_{c} = \sqrt{\frac{2}{\Gamma_c + 1}} a_{c}. 
\label{dc.eq}
\ee
The subscript $c$ denotes the quantities at the critical point during accretion flow. At a critical point, 
$d\vel/dx$ is obtained by using the l'Hospital rule, 
\be
\left(\frac{d \vel}{dx}\right)_{x_{c}} = \frac {(d{\cal N}/dx)_{x_c}}{(d{\cal D}/dx)_{x_c}},
\label{dvc.eq}
\ee
where, ${\cal N}$ and ${\cal D}$ are the numerator and denominator of equation (\ref{dv.eq}). The equation becomes,
\be
A {{\left(\frac{d\vel}{dx}\right)}^{2}}_{x_{c}} + B \left(\frac{d\vel}{dx}\right)_{x_{c}} + C = 0,
\label{qdvc.eq}
\ee
where,
$$
A = 2\left[1+\frac{C_p}{\Gamma_c(2N_c+1)}\right],
$$
$$
 B = \frac{2 \vel_c C_p(5x_c-3)}{\Gamma_c(2N_c+1)x_c(x_c-1)}, 
$$
$$
C = \frac{\vel_c^2 C_p(5x_c-3)^2}{2\Gamma_c(2N_c+1)x_c^2(x_c-1)^2}+\frac{\vel_c^2(5x_c^2-6x_c+3)}{(2x_c^2(x_c-1)^2)}+
\frac{3\lambda_c^2}{x_c^4}-\frac{1}{(x_c-1)^3}
$$ and
$$
C_p = \Gamma_c+\frac{\Theta_c}{\Gamma_c+1}\left(d\Gamma/d\Theta\right)_c.
$$

Since eq.(\ref{qdvc.eq}) has two roots, they  may be either real or complex conjugates depending on the discriminant $D_r=B^2-4AC$ of 
this equation. This also gives the nature of the critical points. If $D_r < 0$ then the critical point will be 
spiral or $O-$ type and if $D_r > 0$ then it will either be saddle type (\ie $X$ type if $C/A<0$),
or nodal type (if $C/A>0$). Depending on the relative strengths of thermal energy and rotational energy of the flow,  
the accreting flow may have one to three sonic points. Out of the possible three sonic points, only two are physical (in the sense,
the flow actually passes through them) X-type and one is unphysical O-type which forms in between the two X-type sonic points.
The sonic point which forms closer to the horizon is called the inner sonic point ($x_{ci}$), and the one far away is called
outer sonic point ($x_{co}$). For flows with very low angular momentum or $\lambda$, generally only $x_{co}$
forms, and flows with very high $\lambda$ only $x_{ci}$ forms, but for intermediate $\lambda$, multiple sonic may form
\citep{lt80,f87,c89,cc11,kc13}.

\subsubsection{Shock conditions}

In the domain of multiple sonic points, supersonic matter through outer sonic point, may be slowed down due to the centrifugal 
barrier at $x\lsim$few$\times10~r_g$. If the barrier is strong enough it may produce a centrifugal barrier mediated shock transition. 
In presence of mass loss, the mass and the momentum flux condition across the shock is given by \citep{cd07,kc13},
\be 
\dot{M}_+= \dot{M}_-(1-R_{\dot{m}}),
\label{mfs.eq}
\ee
and,
\be 
W_+ +\Sigma_+\vel_+^2=W_- +\Sigma_-\vel_-^2.
\label{momf.eq}
\ee
Here, subscripts minus (-) and plus (+) denote the quantities of subsonic and supersonic branch of the accretion flows, respectively,
and, if the mass outflow rate is ${\dot M}_{out}$, then the relative mass outflow rate is $R_{\dot{m}}= \dot{M}_{out}/\dot{M}_-$.
Moreover, $W=2pH$ is the vertically integrated thermal pressure. The third condition will define the type of shock. 
If there is no energy loss at the shock front, then we have an adiabatic shock.
\be 
\cal{E}_+= \cal{E}_-.
\label{enf.eq}
\ee
If there is some energy exchange, it becomes a dissipative shock. In this case,
\be
{\cal E}_+={\cal E}_--\Delta{\cal E}, \mbox{ where, } \Delta{\cal E}=f_e(h_+-h_-),
\label{isoth.eq}
\ee
here, $f_e$ is the dissipation parameter.
If the entropy is conserved across the shock front, then it is called an isentropic shock, and is given by,
\be
{\cal K}_+={\cal K}_-.
\label{isen.eq}
\ee

Using the adiabatic shock conditions \ie (\ref{mfs.eq}-\ref{enf.eq}), the supersonic branch temperature 
and bulk velocity can be obtained from post-shock quantities and vice versa,
\be
\vel_-^2-2(c_1-h_-)=0 \\~~~~~~\mbox{and} \\~~~~~~\Theta_-=\frac{K}{2}(c_0 \vel_--\vel_-^2),
\label{suq.eq}
\ee
here, $c_0=(1-R_{\dot{m}})[2\Theta_+/K+\vel_+^2]/\vel_+$ and $c_1=\vel_+^2/2+h_+$. 
Both the expressions ($\vel_-, \Theta_-$) in eq. \ref{suq.eq} are obtained simultaneously in terms of post-shock quantities
which gives us the shock location $x_s$.

In case of dissipative shocks, we use eqs.(\ref{mfs.eq}, \ref{momf.eq}, \ref{isoth.eq}) to relates post-shock and pre-shock
quantities. The relation is,
\be
\vel_-=\sqrt{2(c_{d}-h_-(1+f_e))}, \mbox{ and, } \Theta_-=\frac{K}{2}(c_0 \vel_--\vel_-^2),
\label{iso.eq}
\ee
where, $c_d=\vel^2_{+}/2+h_+(1+f_e)$. 

For isentropic shocks, we use eqs. (\ref{mfs.eq}, \ref{momf.eq}, \ref{isen.eq}) to obtain,
\be
\vel_- \mbox{exp}(k_{3-})\Theta_-^2(3\Theta_-+2)^{k_{1}}(3\Theta_-+2/\eta)^{k_{2}}-c_{e0}=0; \mbox{ and, }
\Theta_-=\frac{K}{2}(c_0 \vel_--\vel_-^2),
\label{ise.eq}
\ee
where, $c_{e0}=\vel_+ \mbox{exp}(k_{3+})\Theta_+^2(3\Theta_++2)^{k_{1}}(3\Theta_++2/\eta)^{k_{2}}$. 
We will briefly show the three shock transitions in the ${\cal E}-{\md}$ parameter space,
and the related solutions (see Fig. 4). However, in the rest of the paper, we will concentrate on
adiabatic shocks and the resulting mass outflow rate.
 
\subsubsection {Outflow}
Simulations \citep{mrc96,gc13} have shown that extra thermal gradient force along the vertical direction,
drives bipolar outflows, therefore the outflow and the accretion solution is connected by the shock.
As has been suggested by \citet{mrc96} 
and is described in details by \citet{kc13}, 
the outflowing matter tend to move out through the region between the funnel wall (FW) and 
the centrifugal barrier (CB). The cylindrical radius of the CB is, 
\be
x_{CB}=[2 \lambda^{2}r_{CB}(r_{CB}-1)^{2}]^{\frac{1}{4}}, 
\label{cb.eq}
\ee
and that of FW is,
\be
x_{FW}^{2}=\lambda^{2} \frac{(\lambda^{2}-2)+\sqrt{(\lambda^{2}-2)-4(1-y_{CB}^{2})}}{2},
\label{fw.eq}
\ee
where, the spherical radius of CB is $r_{CB}=\left(x_{CB}^{2}+y_{CB}^{2}\right)^{1/2}$, while $y_{CB}$ 
is the local height of CB. The streamline is computed as, $r_{j}=\sqrt{x_{j}^{2}+y_{j}^{2}}$,
where $x_j=(x_{FW}+x_{CB})/2$, and $y_{j}=y_{FW}=y_{CB}$.  The cross-sectional area of the jet is defined as,
\be
{\cal{A}}=\frac{2 \pi (x_{CB}^{2}-x_{FW}^{2})}{\sqrt{1+(dx_j/dy_j)^2}}.
\label{ja.eq}
\ee
The momentum balance equation along the jet stream line is given by
\be 
\vel_j \frac{d\vel_j}{dr} + \frac{1}{\rho_j} \frac{dp_j}{dr} -\frac{\lambda_j^2}{x_j^3}\frac{dx_j}{dr} 
+ \frac{1}{2(r_j-1)^2}\frac{dr_j}{dr} = 0,
\label{jmb.eq}
\ee
here the derivative is with respect to $r=r_{CB}$. Suffix `$j$' denotes jet quantities.
The integrated continuity equation along the jet stream line is given by
\be
\dot{M}_{out}=\rho_j \vel_j \cal{A}
\label{mout.eq}
\ee
This integration constant is called as mass flux which is moving along the stream line of jet.
By using eqs. (\ref{aee.eq}, \ref{mout.eq}), we can write entropy-outflow rate of the jet.
\be
{\dot {\cal M}}_{out}=\frac{\dot{M}_{out}}{2\pi{\cal K}}=\mbox{exp}(k_3) \Theta^{3/2}_j(3\Theta_j+2)^{k_1}
(3\Theta_j+2/\eta)^{k_2}\vel_j
\frac{\cal A}{2\pi}.
\label{scripto_md.eq}
\ee
Now, using eqs.(\ref{eed.eq},\ref{mout.eq}) in eq.(\ref{cee.eq}), we get
\be
\frac{d\Theta_j}{dr}=-\frac{\Theta_j}{N}\left[\frac{1}{\vel_j}\frac{d\vel_j}{dr}+
\frac{1}{\cal{A}}\frac{d\cal{A}}{dr} \right]
\label{dthj.eq}
\ee
The jet velocity gradient equation obtain by combining eqs.(\ref{enth.eq}, \ref{jmb.eq}-\ref{dthj.eq}), we get
\be
\frac{d\vel_j}{dr}=\frac{N_j}{D_j}.
\label{dvj.eq}
\ee
where, 
$$N_j=\frac{a_j^2}{\cal{A}}\frac{d\cal{A}}{dr}+\frac{\lambda_j^2}{x_j^3}\frac{dx_j}{dr}-\frac{1}{2(r_j-1)^2}
\frac{dr_j}{dr}
$$
and
$$
D_j=\vel_j-\frac{a_j^2}{\vel_j}.
$$
The critical point conditions are,
\be
a_{jc}^2=\left[\frac{1}{{\cal{A}}_c} \frac{d{\cal{A}}_c}{dr} \right]^{-1} \left[\frac{1}{2(r_{jc}-1)^2}
\frac{dr_j}{dr}-\frac{\lambda_j^2}{x_{jc}^3}\frac{dx_j}{dr} \right] ~~~~~~~~~~
\& ~~~~~~~~~~  \vel_{jc}=a_{jc}.  
\label{jcrpt.eq}
\ee
The derivatives at $r_{jc}$ is obtained as,
\be
\left(\frac{d\vel_j}{dr}\right)_{r_c}=\left(\frac{dN_j/dr}{dD_j/dr}\right)_{r_c}; ~~ \left(\frac{d\Theta_j}{dr}\right)_{r_c}=
-\frac{\Theta_{jc}}{N_c}\left[\frac{1}{\vel_{jc}}\left(\frac{d\vel_j}{dr}\right)_{r_c}+
\frac{1}{{\cal A}_c}\left(\frac{d\cal{A}}{dr}\right)_{r_c} \right]
\label{dvjc.eq}
\ee
The subscript $c$ denotes the quantities at critical point in the outflow. 

So, the mass outflow rate in term of flow variables at the shock front is given by,
\be 
R_{\dot{m}}= \frac {\dot M_{out}} {\dot M_{in}}=\frac{1}{[2\pi \Sigma_+\vel_+
r_s/(\rho_j\vel_{j}(r_{cb}){\cal{A}}(r_{cb}))+ 1]} 
\label{rmd.eq}
\ee
where, $\rho_j=\rho(r_b)exp(-x_j/h(r_b))$, $\vel_j(r_{cb})$ and ${\cal{A}}(r_{cb})$ are jet base density,
base velocity and base cross section, respectively. Here, $r_b=(x_{ci}+x_s)/3$ and $r_{CB}=r=\sqrt{r_b^2+h_b^2}$,
$h_b$ being the height of the disc at $r_b$. The angular momentum of the jet $\lambda_j=\lambda|_{(x=r_b)}=\lambda$,
\ie the jets are launched with the angular momentum of the accretion disc at the base of the jet,
and since presently we assume inviscid
flow, the jet and the accretion disc has the same angular momentum.

\subsubsection{Solution Procedure}

The solution procedure is the following, given ${\cal E},\lambda,\xi$ the sonic points of the accretion discs are
computed. If there is only one sonic point, one integrates eqs. (\ref{dth.eq},\ref{dv.eq}) inwards and outwards 
starting from the sonic point and using the sonic point conditions (eqs. \ref{nc.eq}-\ref{qdvc.eq}). 
In the multiple critical (or sonic) point regime \ie MCP regime, we may start with either $x_{ci}$ or $x_{co}$ and solve
eqs. (\ref{dth.eq}-\ref{qdvc.eq}) to obtain the solutions. In case the flow parameters ${\cal E},\lambda,\xi$ fall in the
MCP domain, we also check for the shock conditions (eqs. \ref{mfs.eq}-\ref{enf.eq}) as we integrate the equations 
of motion. In the first iteration we assume $R_{\dot m}=0$. If the shock conditions are satisfied, 
then the solution will jump from the supersonic branch to the 
subsonic branch. In this paper, we have always started our integration from $x_{ci}$ if the parameters 
belonged to the MCP domain, and therefore by locating shock through 
eqs. (\ref{mfs.eq}-\ref{enf.eq}) or equivalently eq. (\ref{suq.eq}), the solution will jump to the supersonic branch
through $x_{co}$. Once the shock is found, we solve eqs. (\ref{dvj.eq}-\ref{dthj.eq}) to find the jet solutions and
the corresponding $R_{\dot m}$ (eq. \ref{rmd.eq}). This value of $R_{\dot m}$ is supplied to eq. (\ref{suq.eq})
and the whole solution procedure is repeated to obtain another shock location. This is repeated till the shock 
location converges.  

\section{Results}
\begin{figure}
 \begin{center}
  \includegraphics[width=12.cm]{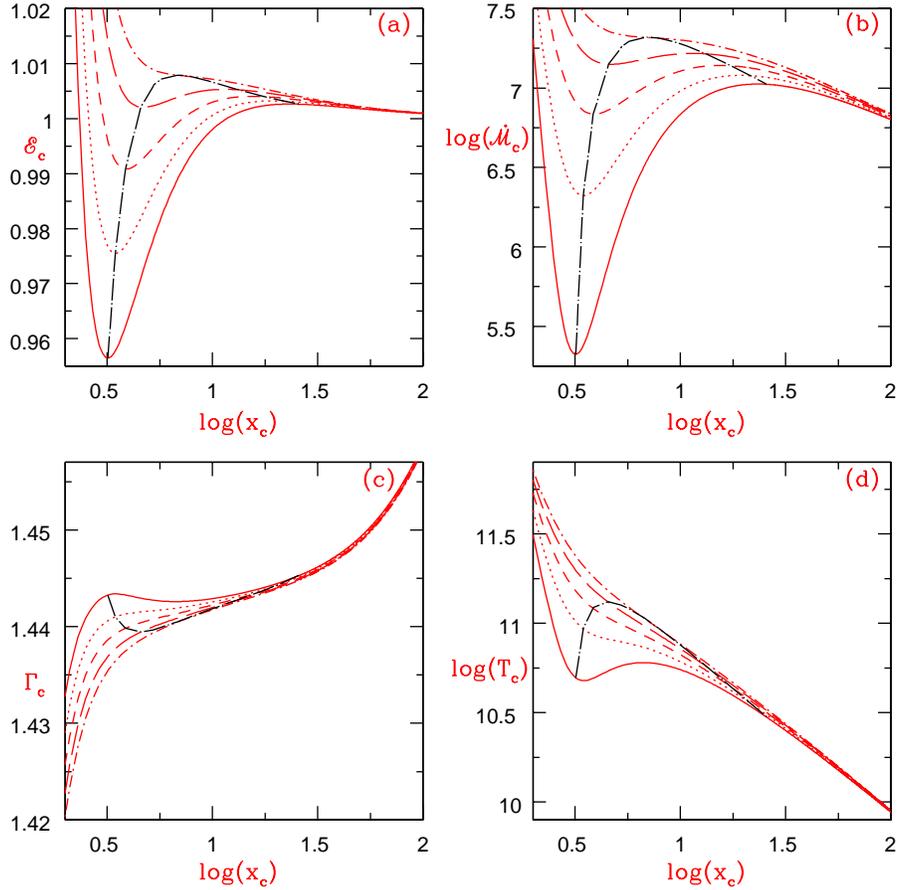}
\caption{(a) $\eng_c$, the specific energy at $x_c$, (b) entropy accretion rate ($\md_c$) at $x_c$, (c) $\Gamma_c$
and (d) $log(T_c)$ is plotted with $log(x_c)$ for $\xi=1$ and for
$\lambda=1.75$ (solid), $1.65$ (dotted), $1.55$ (dashed), $1.45$ (long-dashed), and $1.35825$ (dashed-dotted). 
Long dashed-dotted curve is the loci of maxima and minima of the curves and represents 
the part of the parameter space which gives multiple critical points(MCP).}
 \end{center}
\label{lab:ecencrc}
\end{figure}

\subsection{Nature of critical points and accretion flow solutions}

We discuss the properties of the sonic points of the flow around blackholes with an approximate relativistic
equation of state and how 
they are affected by $\xi$ and $\lambda$ parameters. We plot ${\cal E}_c$ (Fig. 1a), 
${\dot {\cal M}}_c$ (Fig. 1b), $\Gamma_c$ (Fig. 1c) and $log(T_c)$ (Fig. 1d) with $log(x_c)$ for $\xi=1$ 
or electron-proton flow (hereafter $e^--p^+$), and for $\lambda=1.75$ (solid), $1.65$ (dotted), $1.55$ (dashed), 
$1.45$ (long-dashed), and $1.35825$ (dashed-dotted). For low $\lambda$ ($<1.35825$) there is only one 
sonic point for any value of ${\cal E}$, however, the number of sonic points increase with increasing $\lambda$.
Moreover, $\Gamma_c$ depend on the location of the sonic point. Lower temperature at the sonic point or $T_c$
implies lower ${\cal E}_c$, ${\dot {\cal M}_c}$ and higher $\Gamma_c$. Since $\Gamma\rightarrow 5/3$ means non-relativistic
thermal energy and $\Gamma\rightarrow 4/3$ implies ultra-relativistic thermal energy, so higher temperature implies
lower $\Gamma$. The extrema of ${\cal E}_c(x_c)$ and ${\dot {\cal M}}_c(x_c)$ are joined by the long dashed-dotted curve in
the figures which shows the domain of the parameters which supports multiple critical points (MCP).

\begin{figure}
 \begin{center}
  \includegraphics[width=12.cm]{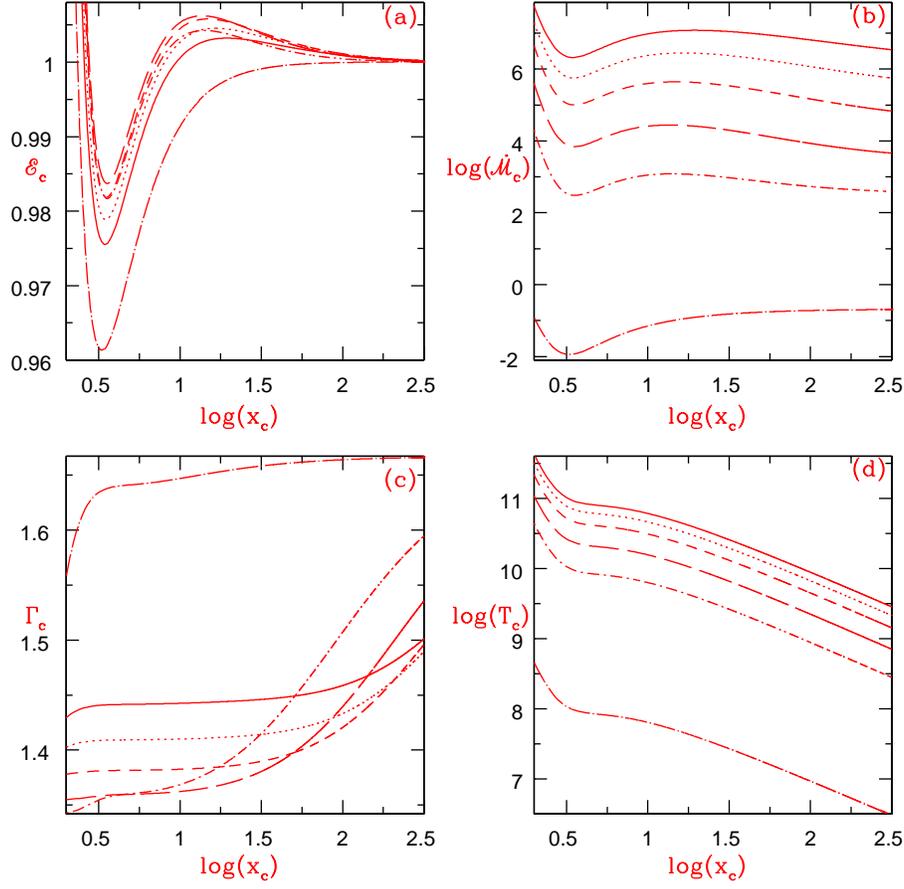}
\caption{(a) $\eng_c$, the specific energy at $x_c$, (b) entropy accretion rate ($\md_c$) at $x_c$, (c) $\Gamma_c$
and (d) $log(T_c)$ is plotted with 
$log(x_c)$ for $\lambda=1.65$ and for
$\xi=1.0$ (solid), $0.75$ (dotted), $0.5$ (dashed), $0.25$ (long-dashed), $0.1$ (dashed-dotted) and $0.0$
(long dashed-dotted).}
 \end{center}
\label{lab:ecxi} 
\end{figure}

To see how the sonic point or the critical point properties are affected by $\xi$, we plot
${\cal E}_c$ (Fig. 2a), ${\dot {\cal M}}_c$ (Fig. 2b),
$\Gamma_c$ (Fig. 2c) and $log(T_c)$ (Fig. 2d) with $log(x_c)$, for $\lambda=1.65$,
and $\xi= 1.0$ (solid), $0.75$ (dotted), $0.5$ (dashed), $0.25$ (long-dashed), $0.1$ (dashed-dotted) and $0.0$
(long dashed-dotted). For flows with $1.0\geq \xi > 0.0$ a maxima and a minima in ${\cal E}_c$ and ${\dot {\cal M}_c}$
exist, however for $\xi=0.0$ there is only a minima and no maxima. Closer inspection reveals that
for $\xi=0$ or pair plasma flow or $e^--e^+$ flow, the physical X type sonic points are formed closer to the horizon,
which means, no matter what the value of $\lambda$, only one sonic point will form for $e^--e^+$ flow, or for
any flow described by
single species EoS. As we have mentioned in \S 2.2.2, that sonic points can only be harboured by hot flows,
and Fig. 2d shows that $T_c$ is the lowest for $e^--e^+$ flow (long dashed-dotted). A flow is thermally relativistic if
its thermal energy is comparable to its rest energy \ie $\Theta=kT/m_ec^2 \gsim 1$, and the adiabatic index will be reflected
by a value
$4/3 \leq \Gamma < 5/3$. Although, $e^--e^+$ is the lightest flow, its $T_c$ is so low that for most values
of $x_c$, $\Gamma_c \sim 5/3$ (long dashed-dotted in Fig. 2c). Therefore, at a large distance away from the black hole,
physical sonic points are not formed. Closer to the horizon, the flow is hot enough to produce
one physical sonic point. As $\xi$ increases, the rest energy increases but
the temperature increases even more, which makes flow with protons much hotter 
and thermally more relativistic (Fig. 2c). Therefore, for a given 
${\cal E}>1$, one or more sonic points may form.

\begin{figure}
 \begin{center}
  \includegraphics[width=12.cm]{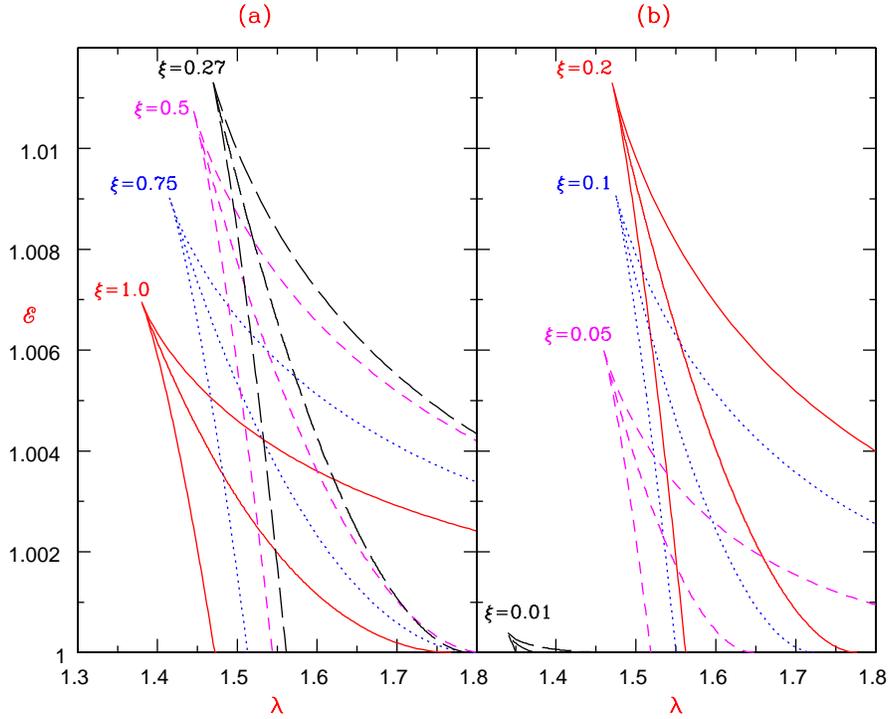}
\caption{$\eng - \lambda$ parameter space represents multiple critical points (MCP) domain for 
(a) $\xi=1.0$ (solid), $\xi=0.75$ (dotted), $\xi=0.5$ (dashed) and $\xi=0.27$ (long-dashed)  and 
(b) $\xi=0.2$(solid), $\xi=0.1$ (dotted), $\xi=0.05$ (dashed) and $\xi=0.01$ (long-dashed).
}
 \end{center}
\label{lab:elmcp}
\end{figure}

One may join the maxima and minima of the ${\cal E}_c(x_c)$ and the corresponding $\lambda$-s for a given $\xi$, and plot
in ${\cal E}-\lambda$ plane, the resulting bounded region gives the values of ${\cal E},~\lambda$ which supports multiple
sonic points, and are abbreviated as MCP region (multiple critical point). In Fig. 3a, we plot the MCP for
$\xi=1.0$ or $e^--p^+$ (solid), $\xi=0.75$ (dotted), $\xi=0.5$(dashed) and $\xi=0.27$(long-dashed), while
in Fig. 3b, we plot $\xi=0.2$ (solid), $\xi=0.1$ (dotted), $\xi=0.05$ (dashed) and $\xi=0.01$ (long-dashed).
It shows that as $\xi$ is reduced, the reduction of rest energy is making the flow more energetic and relativistic,
so the MCP shifts to higher energy, higher angular momentum side. This continues till $\xi=0.27$. Any further reduction
of $\xi$, reduces the temperature to the extent that simultaneous reduction in rest energy cannot compensate
for the lack of thermal energy, and the flow becomes
less energetic and less relativistic, compared to flow of $\xi=0.27$. 
The MCP shifts to the less energetic and  lower angular momentum part of the parameter
space, with simultaneous reduction in the area of the MCP which finally vanishes for $\xi=0.0$.

\begin{figure}
 \begin{center}
  \includegraphics[width=10.cm]{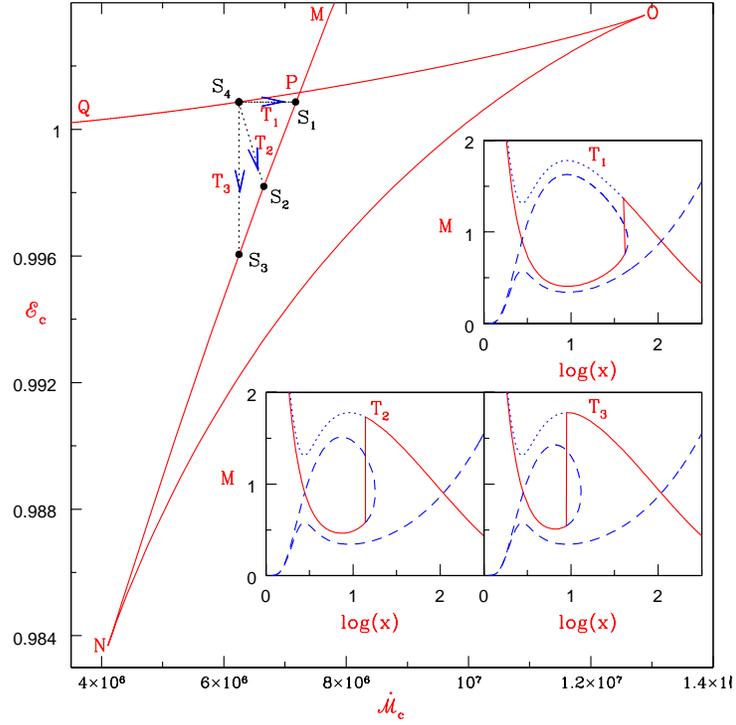}
\caption{Plot of $\eng_c$ with $\md_c$ 
for $\lambda=1.6$ and $\xi=1.0$. Branches $MN, NO$ and $OQ$ corresponds to inner,
 middle and outer critical points of the flow, respectively.
The adiabatic shock transition $T_1$ shown as $S_4\rightarrow S_1$, 
the dissipative shock $T_2$, shown as $S_4 \rightarrow S_2$ and isentropic shock $T_3$, shown as $S_4 \rightarrow
S_3$.
The solutions corresponding to transitions $T_1, T_2, T_3$ are shown in the inset, where solid
curves represent accretion flows with shocks. The coordinates of the transitions are
$S_1(\eng=1.0087, \md=7.17\times 10^6), S_2(\eng=0.9982, \md=6.65\times10^6)$, 
$S_3(\eng=0.99605, \md=6.25\times10^6)$, and $S_4(\eng=1.0087, \md=6.25\times10^6)$.}
 \end{center}
\label{lab:ecenca}
\end{figure}

The sonic point properties not only tell us about the number of sonic points, but also tells us about
the very nature of transitions in the solution. In Fig. 4, we plot $\eng_c$ with $\md_c$ 
for $\lambda=1.6$ and $\xi=1.0$. The inner sonic point quantities are plotted along $MN$, while
middle and outer sonic point quantities are $NO$ and $OQ$, respectively. Adiabatic shocks occur parallel
to the $\md_c$ axis, \ie along $S_4 \rightarrow S_1$, the $T_1$ inset shows the actual Mach number ($M=v/a$) solution
of such a transition (Eq. \ref{suq.eq}). Equation \ref{iso.eq} represents the dissipative shock shown by $S_4 \rightarrow
S_2$ or $T_2$ transition. An isentropic shock is parallel to the ordinate represented by eq. \ref{ise.eq},
and by the transition $S_4 \rightarrow S_3$ or $T_3$ transition. The energy and entropy jumps for the respective transitions
are marked in the Figure. We plot $\eng_c$ with $\md_c$ 
for $\lambda=1.6$ and $\xi=1.0$ (Fig. 5a), $\xi=0.1$ (Fig. 5b), $\xi=0.035$ (Fig. 5c) and $\xi=0.0$ (Fig. 5d).
For a flow with a significant proton proportion, all the branches for multispecies flow are obtained. However,
as the proton proportion is decreased, the outer critical point branch starts to get shortened,
and actually disappears for $\xi=0.0$. This implies there cannot be any steady state shock transitions
for $\xi=0.0$.

\begin{figure}
 \begin{center}
  \includegraphics[width=10.cm]{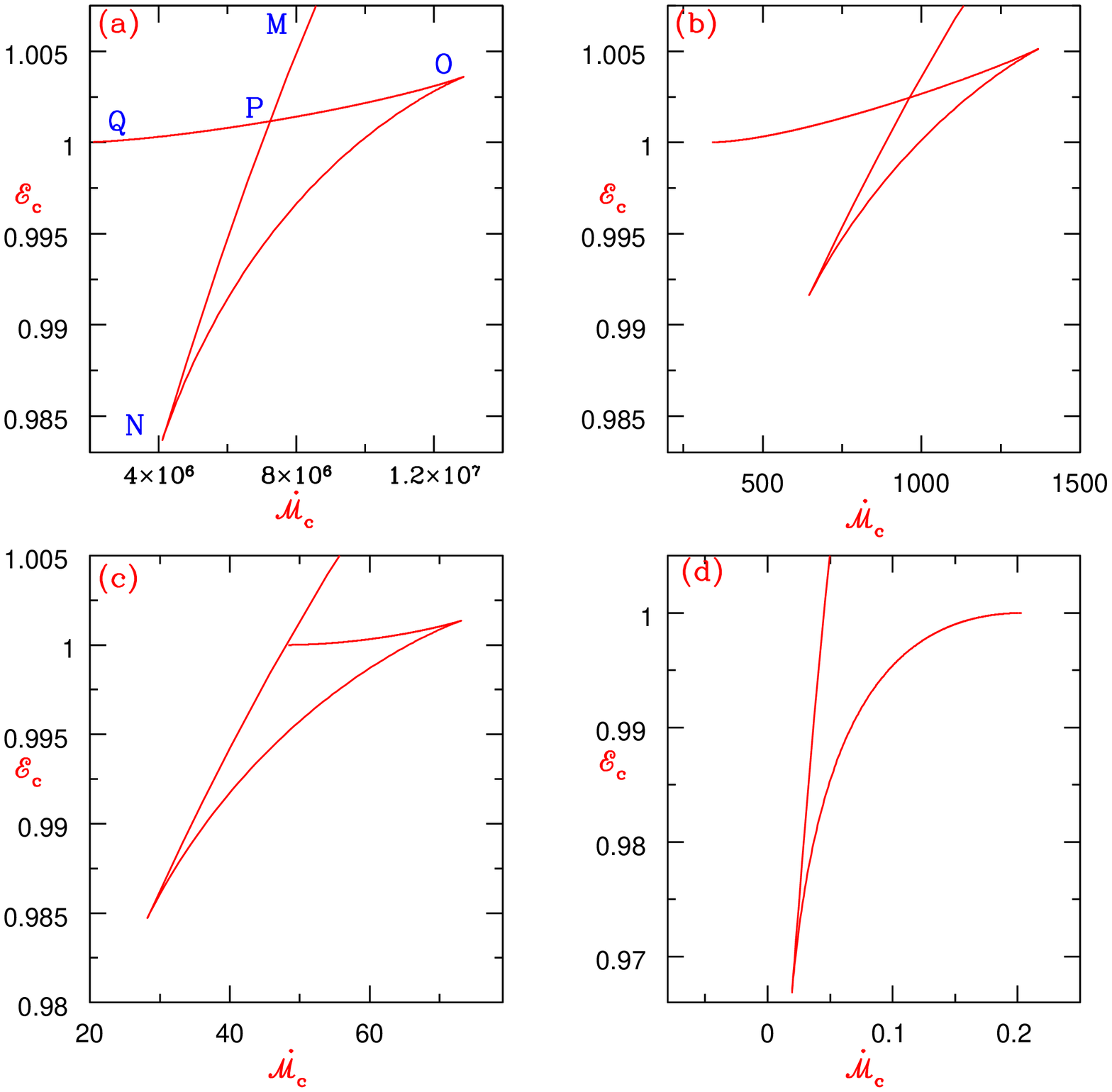}
\caption{$\eng_c$ is plotted with $\md_c$
for $\lambda=1.6$ for various (a) $\xi=1.0$, (b) $\xi=0.1$, (c) $\xi=0.035$, and (d) $\xi=0.0$.
 }
 \end{center}
\label{lab:ecenc}
\end{figure}

\begin{figure}
 \begin{center}
  \includegraphics[width=12.cm]{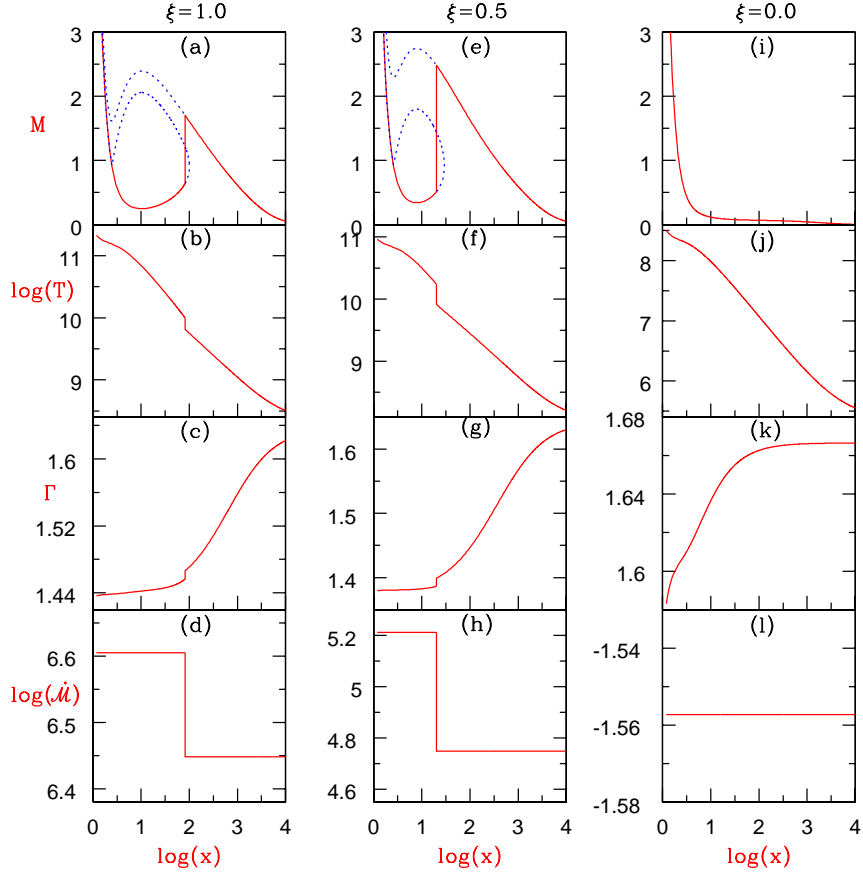}
\caption{Various accretion flow quantities are plotted with radial distance [$log(x)$]: 
(a, e, i) Mach number ($M$), (b, f, j) temperature ($T$), (c, g, k) adiabatic index ($\Gamma$) and 
(e, h, l) entropy accretion rate ($\md$) for the accretion disc parameters, ($\eng,\lambda)=(1.0001, 1.68)$ with 
$\xi=1.0$ (a-d), $0.5$ (e-h) and $0.0$ (i-l).}
 \end{center}
\label{lab:solxi1.5.0}
\end{figure}

Various flow quantities of only the accretion disc (\ie without allowing for mass loss) are plotted in Figs. 6a-l.
The flow variables are $M$ (Figs. 6a, e, i), $T$ (Figs. 6b, f, j), $\Gamma$ (Figs. 6c, g, k) and $\md$ (Figs. e, h, l)
as a function of the radial distance [$log(x)$].
From Figs. 3a-b, it can be shown that for a particular ${\cal E}$ \& $\lambda$, if one $\xi$ value
produces multiple critical points, then another $\xi$ can  exhibit only one sonic point.
However, for Figs. 6a-l, we have chosen a value of $\eng~ \& ~\lambda$ which admits shock solutions for a wide range of
$\xi~(=1\rightarrow 0.087)$. In Figs. 6a-l, we consider the same $(\eng,\lambda)=(1.0001, 1.68)$, but
different compositions: $e^--p^+$ or $\xi=1.0$ (in Figs. 6a-d), $\xi=0.5$ (in Figs. 6e-h), and $e^--e^+$ or $\xi=0.0$
(in Figs. 6i-l). 
The temperature and the entropy is higher for higher $\xi$, and the shock location is also at a larger distance for
flow with higher $\xi$.
Figures 6c, g \& k, show $\Gamma$ is variable for flows with any $\xi$. At $x ~ \rightarrow$ large, $\Gamma \sim 5/3$,
irrespective of the value of $\xi$.
However, since $\xi=0.0$ flow has very low thermal energy, $\Gamma \sim 5/3$ up to $x\sim 100$. But for flows with $\xi \neq
0.0$, we find $\Gamma_{\xi=0.5}<\Gamma_{\xi=1.0}$ at $x<$few$\times 100$. At $x\sim 1$, $\Gamma_{\xi=0.5}\sim 1.4$,
$\Gamma_{\xi=1.0}\sim 1.44$.
This means that the flow becomes thermally more relativistic
with the reduction of $\xi$ up to a certain value (\ie around $\xi\sim 0.27$), and then becomes less relativistic
with further reduction of $\xi$. Although $\xi=0.5$ and $e^--p^+$ flow for the chosen value of ${\cal E}$ \& $\lambda$
harbours shocks, but shock location and other flow parameters are quite different. $e^--e^+$ being the coldest, slowest
and showing only a single sonic point, is significantly different
from any flow with $\xi\neq 0.0$.

\begin{figure}
 \begin{center}
  \includegraphics[width=10.cm]{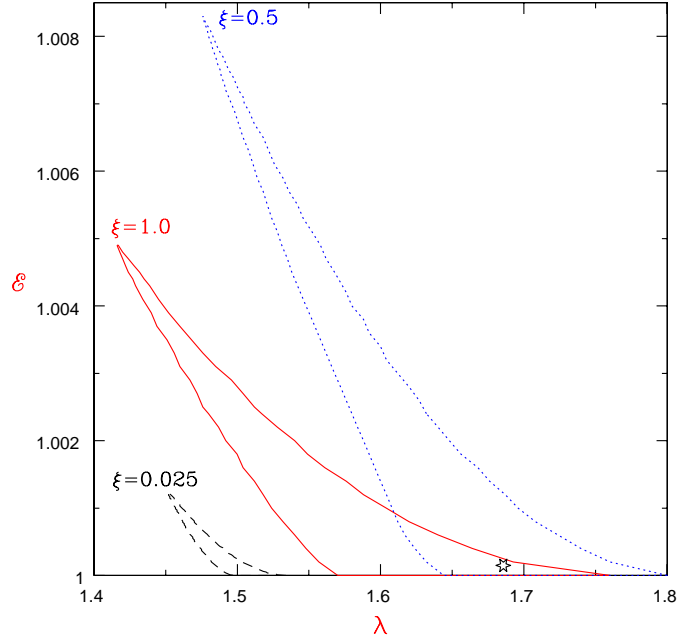}
\caption{Parameter spaces producing shocks without any mass loss:
for $\xi=1.0$ or $e^--p^+$ (solid), $\xi=0.5$ (dotted) and $\xi=0.025$ (dashed).
$e^--e^+$ or $\xi=0.0$ flow does not admit standing accretion shock solution. 
The star marks the parameters with which solutions in Figs. 6a-l, were generated.}
 \end{center}
\label{lab:sokspwom}
\end{figure}

In Fig. 7, we plot the sub-domain of MCP in the ${\cal E}-\lambda$ parameter space which admits standing shock solutions,
for three $\xi$ values, and they are $\xi=1.0$ (solid), $\xi=0.5$ (dotted), and $\xi=0.025$ (dashed).
Similar to Figs. 3a-b, the shock parameter space also shows a shift to the more energetic part of the parameter
space with the decrease of $\xi$. As it has been explained, that the reduction of $\xi$ reduces the rest energy of the flow,
making the flow more relativistic as well as more energetic. But reduction of $\xi$ entails a reduction of $T$ as well,
and for proton poor flows (\ie $\xi \rightarrow 0$) the temperature is so low that it becomes thermally non-relativistic
as well as less energetic. Hence if the shock can at all be supported, it occurs in a low energy and low angular momentum
part of the parameter space. For $\xi=0$ shock in accretion is totally absent and therefore the shock parameter space
for $e^--e^+$ flow does not
appear in the ${\cal E}-\lambda$ space. The star mark indicates the values of ${\cal E}$ \& $\lambda$ which were
used to obtain the solutions of Figs. 6a-l. 

\subsection{Accretion and ejection solutions}

\begin{figure}
 \begin{center}
  \includegraphics[width=10.cm]{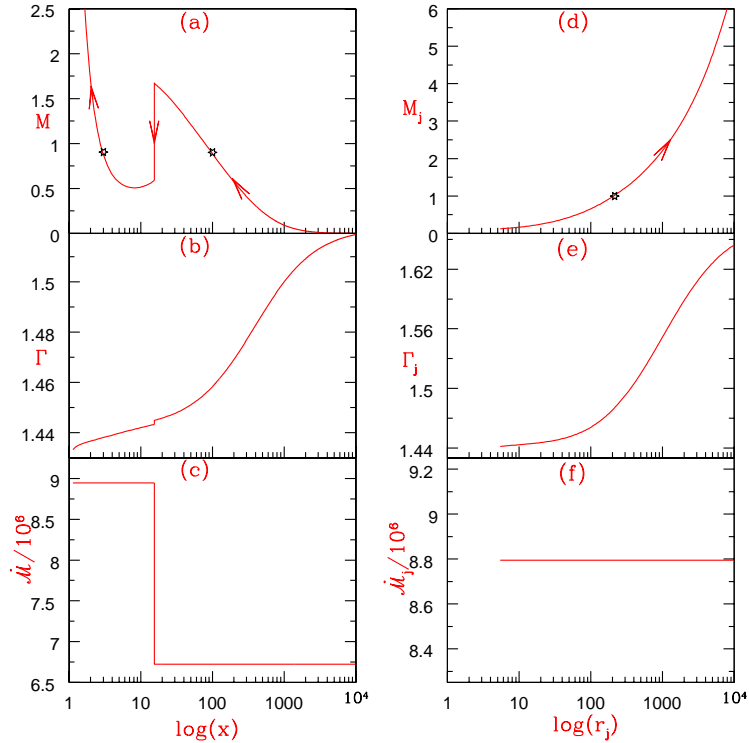}
\caption{Accretion-jet solution for $e^--p^+$ flow, for ${\cal E}=1.001$, $\lambda=1.56$. The accretion shock
denoted by the vertical jump is at $x_s=15.306$. The accretion flow variables $M$ (a), $\Gamma$ (b) and the 
entropy-accretion rate ${\dot {\cal M}}$ (c) are plotted with $x$. The jet variables $M_j=\vel_j/a_j$ (d), $\Gamma_j$ (e), and
${\dot {\cal M}_j}$ (f) plotted with jet radial coordinate $r_j$. Arrows show the direction of the flow and the stars
denote the critical or sonic points of the flow.}
 \end{center}
\label{lab:topxi1}
\end{figure}

In this Section, we present self-consistent accretion-ejection solutions. The methodology to simultaneously and self-
consistently compute the accretion-ejection solution has been presented in Section 2.2.5. In Figs. 8a-c, the accretion disc
flow quantities like $M$ (a), $\Gamma$ (b), and ${\dot {\cal M}}$ (c) are plotted as a function of radial distance.
In Figs. 8d-f, the jet flow quantities \eg $M_j$ (d), $\Gamma_j$ (e), and ${\dot {\cal M}_j}$ (f) are plotted as a
function of $r_j$. The constituent of the flow of the disc-jet system presented in this Figure is $\xi=1.0$.
Other flow parameters are: ${\cal E}=1.001$, $\lambda=1.56$. The 
accretion disc solution admits an accretion shock at $x_s=15.306$ (vertical jump in Figs. 8a-c), and launches a thermally driven
bipolar jet whose sonic point is at $r_{jc}=209.139$. The relative mass outflow rate is $R_{\dot m}=0.038$.
The jet starts with the $\Gamma$ value of the post-shock disc but since it is thermally
driven and is powered by converting the thermal energy to kinetic energy, $\Gamma$ approaches
non-relativistic values far away from the central object. At the shock, the entropy of the jet jumps up from pre-shock
to post-shock value (Fig. 8c).
Interestingly, the entropy of the jet is also much higher than the pre-shock disc. This entropy condition ensures
that although most of the matter flows through the inner sonic point into the black hole, a significant amount
of matter also flows out as jet.

\begin{figure}
 \begin{center}
  \includegraphics[width=12.cm]{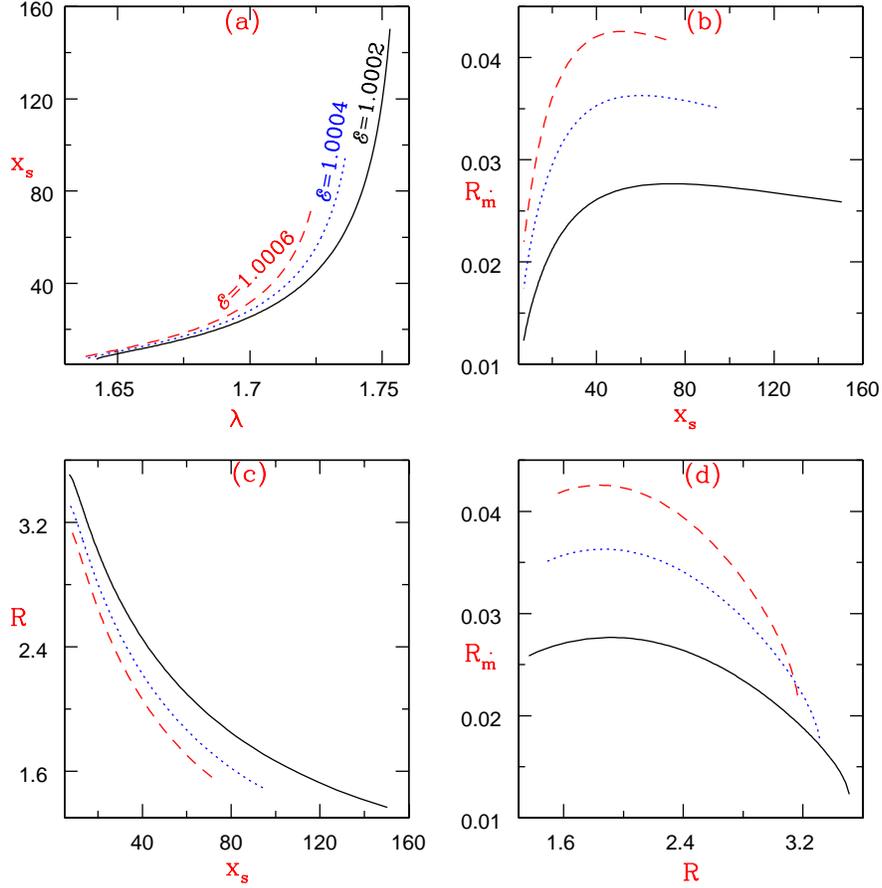}
\caption{
Plot of (a) $x_s$ with $\lambda$, (b) $R_{\dot{m}}$ with $x_s$, (c) Compression ratio
$R$ with $x_s$ and (d) $R_{\dot{m}}$ with $R$. Each curve corresponds to ${\cal E}=1.0002$
(solid), $1.0004$ (dotted), and $1.0006$ (dashed).}
 \end{center}
\label{lab:rmxi1}
\end{figure}
\vskip 2.0cm

Now we plot $x_s$ with $\lambda$ (Fig. 9a), $R_{\dot{m}}$ with $x_s$ (Fig. 9b), compression ratio $R=\Sigma_+/\Sigma_-$
with $x_s$ (Fig. 9c), and $R_{\dot m}$ with $R$ (Fig. 9d), the composition of the flow is given by $\xi=0.27$.
Parameters for each curve is ${\cal E}=1.0002$
(solid), $1.0004$ (dotted), and $1.0006$ (dashed).
For a given value of $\lambda$, $x_s$ increases with ${\cal E}$. Similarly for a given $x_s$, $R_{\dot m}$
increases with ${\cal E}$, but $R$ decreases with the increasing ${\cal E}$. This obviously means $R_{\dot m}$
increases with ${\cal E}$. Interestingly, the dependence of $R_{\dot m}$ with $R$ for a given ${\cal E}$,
qualitatively follows the pattern of \citet{c99}. For a given ${\cal E}$, $x_s$ decreases with decreasing 
$\lambda$. This increases the compression, and drives more matter into the jet channel. But with
decreasing $x_s$, the post-shock area decreases too, and that also limits the total amount of matter
leaving the disc. Hence $R_{\dot m}$ will depend on increasing $R$, as well as the decreasing total
post-shock area, and hence $R_{\dot m}$ peaks at some intermediate value of $R$.

Keeping ${\cal E}=1.0003$, we now vary $\lambda$, of flow with following compositions $\xi=1.0$ (solid),
$\xi=0.635$ (dotted), and $\xi=0.27$ (dashed). We plot $x_s$ with $\lambda$ (Fig. 10a), $R_{\dot m}$ with $x_s$ (Fig. 10b),
$R$ with $x_s$ (Fig.10c), and $R_{\dot m}$ with $R$ (Fig. 10d). We reconfirm that indeed $R_{\dot m}$
increases with the increasing $R$ or decreasing $x_s$, although decreasing post-shock surface area finally decreases
$R_{\dot m}$. $R_{\dot m}$ is highest for $\xi=0.27$, compared to flow of other compositions. To ensure
this we scan the entire
${\cal E}$ \& $\lambda$ parameter space for a given value of $\xi$, and find out the maximum mass outflow rate possible for that
particular $\xi$. In Fig. 11, we plot the maximum mass-outflow rate \ie $R^m_{\dot m}=max(R_{\dot m})$,
as a function of $\xi$. It easily
shows that maximum outflow is possible for $\xi=0.27$.

\begin{figure}
\begin{center}
\includegraphics[width=10.cm]{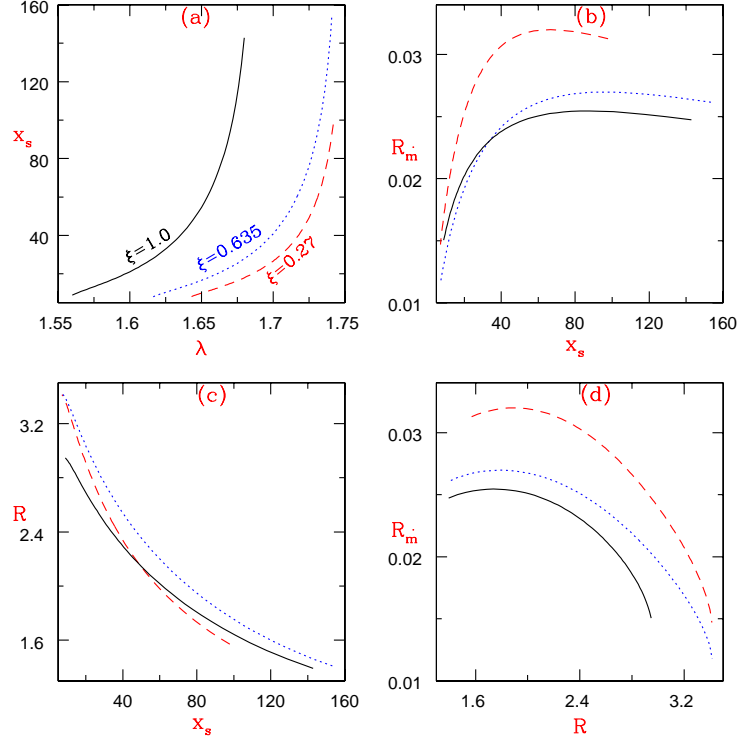}
\caption{
(a) Plot of $x_s$ with $\lambda$, (b) $R_{\dot m}$ with $x_s$, (c) $R$ with $x_s$, and (d) $R_{\dot m}$ with $R$
for ${\cal E}=1.0003$. Each curve corresponds to $\xi=1.0$ (solid), $0.635$ (dotted), and $0.27$ (dashed).}
 \end{center}
\label{lab:rmdxi}
\end{figure}

\begin{figure}
 \begin{center}
  \includegraphics[width=9.cm]{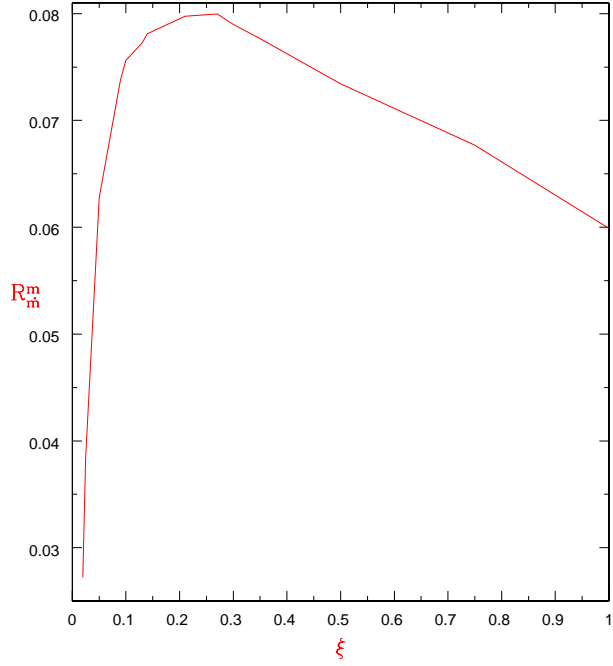}
\caption{The plot of maximum mass outflow rate $R^m_{\dot m}$ with the composition parameter $\xi$.}
 \end{center}
\label{lab:sokspwm}
\end{figure}

\begin{figure}
 \begin{center}
  \includegraphics[width=10.cm]{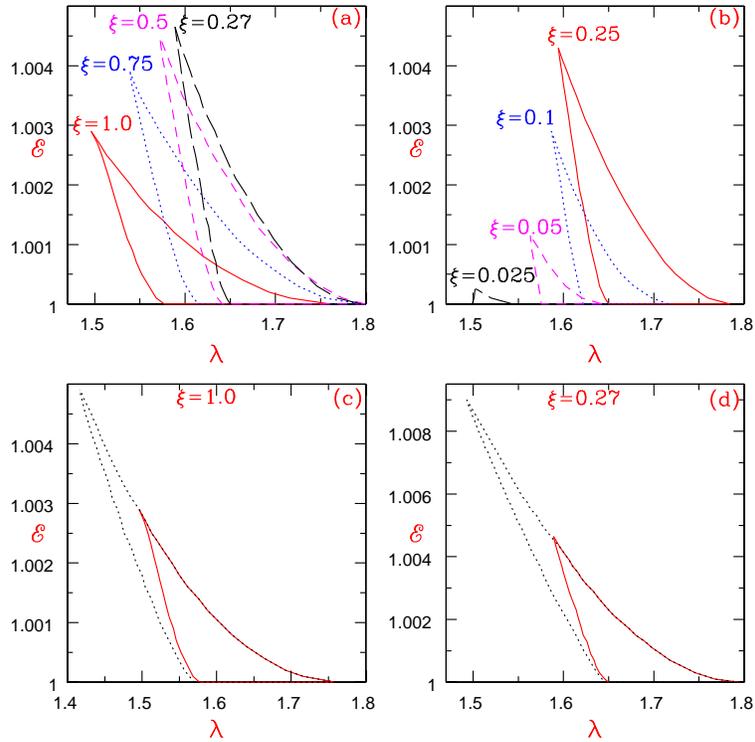}
\caption{Shock-domain in the ${\cal E}-\lambda$ parameter space with mass loss. Flow composition
$\xi$ is marked. (a) Parameter space for $\xi=1.0 \rightarrow 0.27$ and (b) for $\xi=0.25 \rightarrow 0.025$.
(c) Comparison of the domains in ${\cal E}-\lambda$ space with mass loss (solid) and without mass loss (dotted)
for $\xi=1.0$ and (d) Comparison of the 'shock-domain' with mass loss (solid) and without mass loss (dotted)
for $\xi=0.27$. }
 \end{center}
\label{lab:sokspwm}
\end{figure}

In Figs. 12a-b, we show the parameter space where standing shocks form (the so-called `shock-domain') for various $\xi$. 
Similar to the behaviour of MCP domain, the shock domain in ${\cal E}-\lambda$ parameter space shifts to the higher energy
and angular
momentum region as $\xi$
is decreased from its $e^--p^+$ value up to $\xi=0.27$. If $\xi$ is decreased further, the flow becomes less energetic
and the shock-domain moves towards the lower energy and lower angular momentum corner, with subsequent decrease in
the bounded area of the shock space. Finally, the shock domain disappears for $\xi=0.0$.
In Fig. 12c, we compare the shock domain for $\xi=1.0$, of the accretion flow which includes mass loss (solid)
and which does not include mass loss (dotted). In Fig. 12d, we compare the shock domain for $\xi=0.25$, of the accretion
flow which includes
mass loss (solid) and which does not mass loss (dotted). This shows that the reduction of post-shock pressure due to mass
loss,
reduces the steady shock domain. 
Flow from that part of the parameter space which admits shock without massloss,
actually do not show steady shock when massloss is allowed. So there is a possibility of massloss driven 
shock oscillation too.

\section{Discussion and Concluding Remarks}

In this paper, we studied the outflow behaviour taking into account flow composition and approximate relativistic
equation of state of the flow onto black holes. We have re-established our earlier findings that
$e^--e^+$ is the least relativistic flow, but $e^--p^+$ is not the most relativistic flow. Thermally, the most relativistic
flow is when $\xi=0.27$. As has been explained earlier, the flow starting with the same ${\cal E}$ implies a  
flow starting with the same sound speed at infinity. Since the sound speed is a measure of the thermal speed,
if we compare two flows with the same thermal speed, the constituent particles will transfer less average kinetic energy (\ie thermal energy)
in the flow where these particles are lighter. Therefore, the temperature of $e^--e^+$ flow is lesser than the flow containing
both electrons (and/or positrons) and heavier protons. The temperature of $e^--e^+$ is so low that the lower inertia 
of the flow cannot compensate for its lack of thermal energy and therefore thermally it is the least relativistic, while the flow
with some amount of protons are more relativistic.
Although the conclusions are consistent with the previous papers \citep{c08,cr09,cc11}, in this
paper, we derived the analytical form of the adiabatic equation of state for multispecies flow (\eg eq. \ref{aee.eq}).
Using the adiabatic equation of state, we present the form of the entropy-mass flow rate for accretion (eq. \ref{enar.eq}),
as well as for outflow (eq. \ref{scripto_md.eq}).  
Since the $e^--e^+$ flow is not hot enough, they show only one physical sonic point, and has been shown in this paper.
However, the flow with some amount of protons show formation of multiple sonic points and shocks. 

Post-shock disc drives bipolar outflows. The mass outflow rate is calculated self-consistently and is found to be around a few percent.
Interestingly, the mass outflow rate is also generally higher for flows with $\xi\sim 0.27$. Since, the shock
forms at a larger distance for flows with a higher angular momentum, the mass outflow rate is also generally lower
for high angular momentum flow. The mass outflow rate depends on many factors, the shocks location, the angular momentum, energy or composition
of the accretion flow. If the shock location is smaller (say for lower angular momentum), the compression ratio is larger,
and the mass outflow increases. But we have shown that the mass outflow rate do not increase monotonically with the compression ratio, rather 
the mass outflow rate peaks at some intermediate value. We also plotted the parameter space for the shock in presence of the mass loss. Due to the loss of matter,
the post-shock pressure is reduced, and therefore the entire range of parameters for which steady shock condition 
is satisfied (the so-called shock-domain) 
for flow without mass loss, also shrinks. Hence there is a possibility of mass loss induced shock instability
too.  
Since shocks do not form for $\xi=0.0$ or $e^--e^+$ flow, purely $e^--e^+$ jet solutions are not found, although lepton
dominated ($0<\xi<1$) jets are possible. It must be noted that a purely $e^--e^+$ flow is essentially a flow described 
by fully ionized single species EoS. At temperatures $T> 10^5$ the flow is likely to be fully ionized.
A fully ionized flow comprised of similar particles can only be $e^--e^+$ flow. However, at the same time, an accretion disc
from large distances to the horizon and made up of only the pair plasma is not likely to happen in nature.
We compared $e^--e^+$ flow with other flows containing protons, only to show the stark contrast
between a physically plausible flow and a flow which is not viable. Moreover, it is to be noted that
in principle, $\xi$ should be computed from physical processes self-consistently and not supplied as a parameter. In a given
disc, $\xi$
must be a function of the radial distance. We are working on this question and the results will be communicated elsewhere.

\section*{Acknowledgment}

CBS acknowledges the Visiting Students Programme in ARIES, where this work was initiated.

{}

\end{document}